\documentclass[sn-mathphys,Numbered]{sn-jnl}

\usepackage{graphicx}%
\usepackage{multirow}%
\usepackage{amsmath,amssymb,amsfonts}%
\usepackage{amsthm}%
\usepackage{mathrsfs}%
\usepackage[title]{appendix}%
\usepackage{xcolor}%
\usepackage{textcomp}%
\usepackage{manyfoot}%
\usepackage{booktabs}%
\usepackage{algorithm}%
\usepackage{algorithmicx}%
\usepackage{algpseudocode}%
\usepackage{listings}%
\usepackage[authoryear,round]{natbib}
\usepackage{subcaption}
\usepackage{xurl} % Better URL line breaking
\usepackage{tabularx}
\usepackage{booktabs,siunitx}
\begin{document}

\title{Vision Transformer for Multi-Domain Phase Retrieval in Coherent Diffraction Imaging}

\author[1]{\fnm{Jialun} \sur{Liu}}
\author[2]{\fnm{David} \sur{Yang}}
\author[1,2]{\fnm{Ian} \sur{Robinson}}

\affil[1]{London Centre for Nanotechnology, University College London, London WC1E 6BT, United Kingdom}
\affil[2]{Condensed Matter Physics and Materials Science Department, Brookhaven National Laboratory, Upton, NY 11973, USA}

\abstract{
Bragg coherent diffraction imaging (BCDI) phase retrieval becomes rapidly difficult in the strong-phase regime, where a crystal contains distortions beyond half a lattice spacing. An important special case is the phase domain problem, where blocks of a crystal are displaced with sharp jumps at domain walls. The strong-phase, here defined as beyond $\pm \pi/2$, generates split Bragg peaks and dense fringe structure for which classical iterative solvers often stagnate or return different solutions from different initialisations. Here, we introduce an unsupervised Fourier Vision Transformer (Fourier ViT) to solve this block-phase, multi-domain phase-retrieval problem directly from measured 2D Bragg diffraction intensities. Fourier ViT couples reciprocal-space information globally through multiscale Fourier token mixing, while shallow convolutional front and back-ends provide local filtering and reconstruction. We validate the approach on large-scale synthetic datasets of Voronoi multi-domain crystals with strong-phase contrast under realistic noise corruptions, and on experimental diffraction from a $\mathrm{La}_{2-x}\mathrm{Ca}_x\mathrm{MnO}_4$ nanocrystal. Across the regimes considered, Fourier ViT achieves the lowest reciprocal-space mismatch ($\chi^2$) among the compared methods and preserves domain-resolved phase reconstructions for increasing numbers of domains. On experimental data, with the same real-space support, Fourier ViT matches the iterative benchmark $\chi^2$ while improving robustness to random initialisations, yielding a higher success rate of low-$\chi^2$ reconstructions than the complex convolutional neural network baseline.
}


\maketitle
\newpage
\section{Introduction}
Bragg coherent diffraction imaging (BCDI) has transformed the way we characterise nanoscale materials by revealing the internal structure and lattice distortions of single crystals in three dimensions \citep{miao1999extending,RobinsonHarder2009,pfeifer2006three}. This lensless X-ray technique can invert Bragg peaks into real-space by iteratively applying reciprocal and real-space constraints via fast Fourier transforms  \citep{fienup1982phase}. BCDI reconstructions have revealed how edge strain in single Pt nanocatalysts influences methane-oxidation activity \citep{kim2018active}, how nanostrain accumulation drives structural degradation and voltage fade in Li-rich layered oxide cathodes \citep{liu2022origin}, and how electric fields can affect domain wall dynamics in BaTiO$_3$ nanoparticles \citep{liu2025electric}. However, its core challenge remains in phase retrieval: detectors record only diffraction intensities, so the phase required for the real-space reconstruction is missing \citep{sayre1952shannon,fienup1982phase}.

Recovering phase from intensities alone is a non-convex, ill-conditioned inverse problem whose difficulty grows sharply with internal strain and domain complexity. Classical algorithms such as Gerchberg–Saxton error reduction \citep{gerchberg1972practical}, hybrid input–output (HIO) \citep{fienup1982phase}, difference map \citep{elser2003phase} and Relaxed Averaged Alternating Reflections (RAAR) \citep{luke2005relaxed} alternate between real and reciprocal-space, enforcing constraints such as support and measured intensities. For “weak-phase’’ crystals, where phase shifts stay below about $\pi/2$, these methods are often reliable and converge in a few hundred iterations \citep{RobinsonHarder2009}. Once phase shifts exceed this regime, interference between differently phased regions leads to peak splitting and complex fringe patterns, so called the “strong-phase’’ problem \citep{Robinson2020DomainTexture,Wu2021IUCrJ_PhaseDomains, Gao2021PRB_MultiPeak}. The central Bragg maximum can split into multiple lobes, producing separated intensity maxima and dense fringes around the Bragg peak, so the measured diffraction becomes extremely sensitive to the internal phase distribution. In this strong-phase, multi-domain regime, iterative solvers stagnate or converge to different solutions under different random seeds, reflecting an apparent non-uniqueness of the reconstruction \citep{bates1984uniqueness,ulvestad2017defects,carnis2019strain}. This slow and fragile convergence limits routine use of BCDI and makes real-time feedback for in situ or operando experiments difficult, especially at X-ray free-electron laser (XFEL) and modern synchrotron sources \citep{clark2013ultrafast,harder2021deep}.

Under idealised conditions, theoretical work on the uniqueness of the phase problem shows that oversampling in two or more dimensions is sufficient to guarantee a unique solution up to trivial symmetries such as real-space translation, conjugate inversion, a global phase shift and Fresnel propagation \citep{bates1984uniqueness,BruckSodin1979,zhuang2022,Huang2011uniqueness}. In contrast, the one-dimensional problem generally admits many non-equivalent solutions that share the same Fourier magnitudes \citep{BeinertPlonka2015,Bendory2017FourierPhaseRetrieval}. Motivated by this, we focus on a two-dimensional formulation in which each synthetic crystal is treated as a projection of a three-dimensional nanocrystal, thereby preserving the essential Fourier-phase structure of the full 3D problem. Following Bates \citep{bates1984uniqueness}, we consider the 2D and 3D phase problems to be equivalent. 

We focus on the phase-domain limit relevant to multi-domain crystals. In BCDI, the real-space phase $\phi(\mathbf{r})=\mathbf{Q}\!\cdot\!\mathbf{u}(\mathbf{r})$ is the projection of the lattice displacement field $\mathbf{u}(\mathbf{r})$ onto the reciprocal-lattice vector $\mathbf{Q}$ of the measured Bragg peak \citep{pfeifer2006three,RobinsonHarder2009}. We define a phase domain as a connected block-shaped region in which this projected phase is approximately constant, separated from neighbouring regions by sharp phase discontinuities across narrow domain walls. As the number of domains increases, coherent interference between domains produces stronger Bragg-peak splitting and richer fringe structure, and the phase-retrieval landscape becomes more complicated \citep{Robinson2020DomainTexture}. Such phase domain textures occur in real quantum materials, for example, BCDI has been used to image the formation and scaling of low-temperature orthorhombic domains in the cuprate superconductor La$_{1.875}$Ba$_{0.125}$CuO$_4$ \citep{Assefa2020ScalingLTO} and monoclinic domains in Magnetite Fe$_3$O$_4$ \citep{Dong2025SymmetryBreaking}.

In practice, experimental data can be noisy and constraints such as support or non-negativity are only approximate \citep{Fienup1987Reconstruction}. For noisy data the phase problem no longer has a single exact solution. Instead, one obtains a family of near-solutions whose spread reflects both noise and algorithmic trapping \citep{Bates1972}. This has motivated ensemble strategies, in which many independent reconstructions are run and then either selected or averaged according to a figure of merit. Examples include guided HIO procedures \citep{ulvestad2017defects,chen2007} and the common-mode (eigen-solution) analysis \citep{LeeSeung1999NMF} implemented in \textsc{PyNX} , which combines multiple BCDI reconstructions using a free log-likelihood metric \citep{favre_nicolin2020free,favre_nicolin2020pynx}. Alternative projection schemes, including Fourier-weighted projections \citep{GuizarSicairos2008FWP}, improve robustness for truncated diffraction data but remain iterative projection algorithms and can still be sensitive to initialisation. Overall, phase retrieval for multi-domain crystals remains slow, fragile, and difficult to automate.

In parallel, deep learning is increasingly used to solve coherent imaging problems. Supervised neural networks learn the inverse mapping from diffraction intensities to real-space structure, enabling near real-time reconstructions compared with iterative phase retrieval \citep{cherukara2018real,wu2021three,Wu2025Unveiling}. 
CNNs with U-Net architectures have been applied to optical holography \citep{Rivenson2018LSA}, X-ray ptychography \citep{Cherukara2020PtychoNN}, and coherent X-ray diffraction imaging (CXDI) \citep{vu2025pid3net}, including BCDI \citep{Wu2021IUCrJ_PhaseDomains,wu2021three,yu2024ultrafast}. After training, these networks can return reconstructions orders of magnitude faster than classical iterative solvers \citep{cherukara2018real,Cherukara2020PtychoNN,yu2024ultrafast}. The main limitation of supervised approaches is that they learn only what has been given in the training set, and therefore do not guarantee the correct solution for a new object outside the training distribution, particularly when ground-truth labels are unavailable for experimental patterns \citep{cherukara2018real,wu2021three,yao2022autophasenn}. Since label generation and verification are most straightforward in simple regimes, many published demonstrations and benchmarks focus on single- to few-domain systems \citep{wu2021three}.
Unsupervised and physics-aware variants, such as AutoPhaseNN and deep complex-valued CNNs for ultrafast BCDI, embed the forward imaging model directly into the network and loss function, thereby reducing the dependency on labelled training data \citep{yao2022autophasenn,yu2024ultrafast}. This unsupervised setting is particularly natural for BCDI, where ground-truth real-space phases are not available for experimental patterns and where strong-phase, multi-domain structures make the optimisation landscape highly non-convex and sensitive to initialisation \citep{wu2021three,yao2022autophasenn}. Vision transformers (ViTs) have recently been applied in coherent X-ray imaging, for example in ptychography \citep{gan2024ptychodv}, but have not yet been integrated into a fully unsupervised, physics-informed BCDI framework.

Here, we introduce a simple block-phase domain model to access the strong-phase, multi-domain regime, and solve the resulting phase-retrieval problem using an unsupervised Fourier ViT with global reciprocal-space token mixing. The proposed Fourier ViT combines convolutional feature extraction with spectral token mixing, providing global coupling at $\mathcal{O}(N\log N)$ complexity, in contrast to the $\mathcal{O}(N^{2})$ cost of full dot-product self-attention in standard Transformers \citep{Vaswani2017Attention}. Fourier token mixing has been explored across modalities, from unparameterised Fourier mixing (FNet) to learnable global frequency filters (GFNet) and adaptive Fourier operators (AFNO/FNO) \citep{LeeThorp2022FNet,Rao2021GFNet,Guibas2021AFNO,Li2021FNO}. Here, we adapt these ideas to oversampled Bragg diffraction through a multiscale spectral design tailored to the structure of BCDI patterns.

We first validated Fourier ViT on large-scale synthetic datasets of Voronoi multi-domain crystals. Over the range of conditions studied, Fourier ViT achieves the lowest reciprocal-space mismatch ($\chi^{2}$) among the compared methods, while preserving domain-resolved phase reconstructions under realistic noise degradations. We further apply the method to experimental diffraction from $\mathrm{La}_{2-x}\mathrm{Ca}_{x}\mathrm{MnO}_{4}$ (LCMO) \citep{RobinsonPhaseDomains}. Here, Fourier ViT outperforms a C-CNN baseline \citep{yu2024ultrafast} and achieves $\chi^{2}$ values competitive with the iterative benchmark. The run-to-run $\chi^{2}$ distribution is broader across random initialisations, highlighting the multi-basin nature of strong-phase retrieval.

\begin{figure*}[ht]
    \centering
    \includegraphics[width=\textwidth]{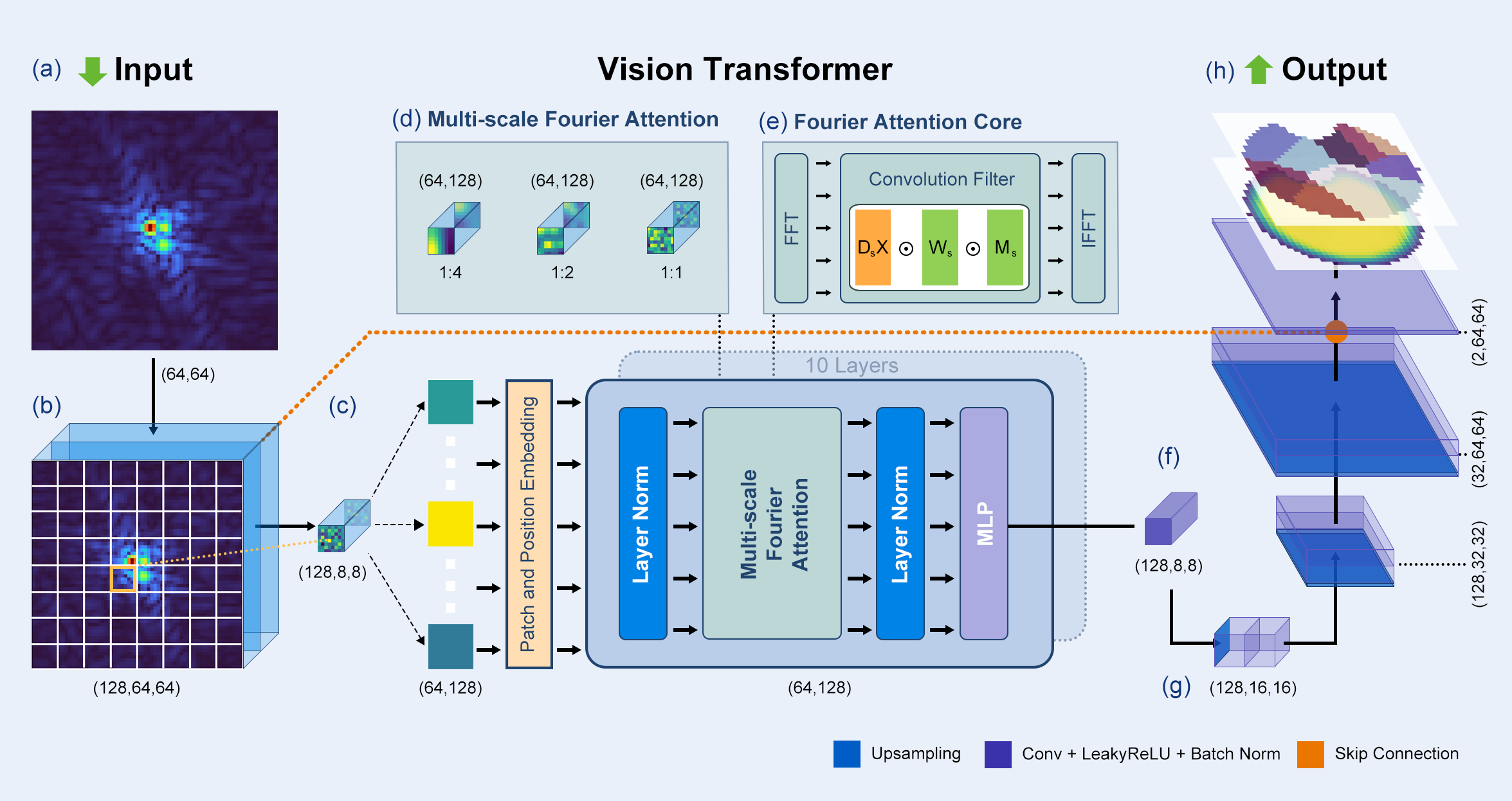}
    \caption{Architecture of the proposed Fourier ViT model for BCDI phase retrieval. (a) The input diffraction magnitude (\(64\times64\) pixels) is passed through (b) a shallow convolutional feature extractor to produce 128 channels (\(128\times64\times64\)). (c) Each image is partitioned into patches and embedded as tokens (\(8\times8\) patches shown for clarity; all experiments use \(16\times16\) patches), giving a token sequence of shape \((64\times128)\). This sequence is processed by (d) a multi-layer Vision Transformer with multi-scale Fourier attention (frequency mixing at 1:4, 1:2 and 1:1 scales) and (e) an FFT-based global convolution block. (f) The transformer output is reshaped into a latent feature map (\(128\times8\times8\)) and decoded by (g) a convolutional upsampling path with a skip connection to reconstruct the complex crystal field, yielding (h) \(64\times64\) real-space amplitude and phase maps consistent with the measured diffraction pattern.}
    \label{fig:cnnvit_architecture}
\end{figure*}

\section{Results}
\subsection{Fourier ViT model architecture}\label{sec:model}
\subsubsection{Tokenisation and positional embedding}\label{sec:token}
As shown by Fig.\ref{fig:cnnvit_architecture} (a), the input is a 2D diffraction pattern magnitude with \(64\times64\) pixels, max-normalised to $[0,1]$. In panel (b), a shallow CNN extracts features at the original $64\times64$ resolution and outputs a 128-channel feature map. We retain this as a full-resolution skip map $\mathbf{S}\in\mathbb{R}^{B\times128\times64\times64}$, where $B$ denotes the batch size. Patch embedding in Fig.~\ref{fig:cnnvit_architecture}(c) uses a single $4\times4$ convolution with stride 4 applied to $\mathbf{S}$, partitioning the image into non-overlapping $4\times 4$ patches and projecting each patch to a token of embedding dimension $E{=}128$. This produces a $16\times16$ token grid, thus $N{=}16\times 16{=}256$ tokens in total. The typical speckle/fringe spacing in our datasets is about two pixels after binning, so a $4\times 4$ patch spans at least one full fringe period and allows each token to encode a physically meaningful local piece of the diffraction pattern. Following the Vision Transformer formulation \citep{Dosovitskiy2021ViT,Vaswani2017Attention}, we then flatten the $16\times 16$ grid into a token sequence and refine it with two $1\times 1$ convolutions before adding a learned positional embedding. The resulting sequence $\mathbf{X}\in\mathbb{R}^{B\times N\times E}$ is passed to the stack of Fourier transformer blocks described in Sec.~\ref{sec:spectral}.

\subsubsection{Multi-scale Fourier attention}\label{sec:spectral}
Inspired by Fourier-based transformers and spectral gating mechanisms \citep{LeeThorp2022FNet,Patro2023SVT}, 
we replace dot product self-attention \citep{Vaswani2017Attention} with a multi-scale Fourier attention, indicated by the green boxes in Fig.\ref{fig:cnnvit_architecture} (d, e). The tokens are reshaped to a feature map $\mathbf{X}\in\mathbb{R}^{B\times E\times H\times W}$ with $H{=}W{=}16$ and processed at three spatial scales $s\in\{1,2,4\}$:
\begin{equation}
\mathbf{Z}_s
= \mathcal{U}_s\!\Big(\mathcal{F}^{-1}\!\big(\,\mathcal{F}(\mathcal{D}_s\mathbf{X})\ \odot\ \mathbf{W}_s\ \odot\ \mathbf{M}_s\,\big)\Big), 
\end{equation}
Here $\mathcal{D}_s$ denotes average pooling with stride $s$, which reduces the spatial size to $h_s{\times}w_s=(H/s){\times}(W/s)$, and $\mathcal{U}_s$ upsamples back to $(H,W)$. The operator $\mathcal{F}$ is the channel-wise 2D FFT, and $\odot$ denotes element-wise multiplication in frequency space. In practice, we use three mixers at resolutions $(H,W)$, $(H/2,W/2)$ and $(H/4,W/4)$ and learn their relative importance via the scale weights $\alpha_s$, giving $\mathbf{Z}=\sum_{s}\alpha_s\,\mathbf{Z}_s$ with $\boldsymbol{\alpha}=\mathrm{softmax}(\boldsymbol{\theta})$. At each scale, we learn per-channel frequency responses $\mathbf{W}_s\in\mathbb{R}^{E\times h_s\times w_s}$ and a channel-shared spectral gate $\mathbf{M}_s\in\mathbb{R}^{1\times h_s\times w_s}$. This explicit multi-scale design allows the block to respond both to finely-spaced fringes and to broader low frequency components.

We use pre-normalised residual connections in each Fourier transformer block:
\begin{equation}
\mathbf{X}'=\mathbf{X}+\mathrm{Mixer}\!\big(\mathrm{LN}(\mathbf{X})\big),\qquad
\mathbf{X}_{\mathrm{out}}=\mathbf{X}'+\mathrm{MLP}\!\big(\mathrm{LN}(\mathbf{X}')\big),
\end{equation}
where $\mathrm{LN}$ represents layer normalisation \citep{Ba2016LayerNorm,Dosovitskiy2021ViT}, $\mathrm{Mixer}$ denotes the multi-scale Fourier spectral mixer defined above, and multilayer perceptron ($\mathrm{MLP}$) is a feed-forward projection with hidden dimension $d_{\mathrm{mlp}}$ and ReLU nonlinearity. The residual paths improve gradient flow, while pre-normalisation stabilises optimisation for deeper stacks of blocks. Additional details, including the convolution theorem and complexity analysis, are given in the Supplementary Note 1.

\subsubsection{Decoder}\label{sec:decoder}
After the final transformer block, the output sequence
$\mathbf{X}\in\mathbb{R}^{B\times N\times E}$ (with $N{=}16\times 16$ and $E{=}128$) is reshaped to an $E\times 16\times 16$ feature map as shown in Fig.\ref{fig:cnnvit_architecture} (f) and upsampled to $64\times 64$ by a small CNN decoder (Fig.\ref{fig:cnnvit_architecture} (g)). At full resolution the decoder fuses three inputs: (i) the upsampled token features, (ii) the encoder skip map $\mathbf{S}\in\mathbb{R}^{B\times 128\times 64\times 64}$, which carries local diffraction pattern context from the CNN feature extractor, and (iii) a frequency–space summary of $\mathbf{S}$ obtained by taking its channel-wise Fourier magnitude, applying $\log\!\bigl(1 + |\mathcal{F}\{\mathbf{S}\}|\bigr)$ to compress the dynamic range, and normalising per channel (see Supplementary Note 2 for details). The three streams are concatenated along the channel dimension and passed through $3\times 3$ convolutions for channel mixing, followed by a short refinement stack. This fusion supplies the decoder with complementary information: (i) non-local diffraction correlations from the transformer, (ii) fine scale speckle features from the encoder, and (iii) an explicit spectral summary of the learned diffraction features.

The decoder heads output a real valued crystal amplitude map $\hat{A}(x,y)$ and two channels $(\hat{c}(x,y),\hat{s}(x,y))$ that parametrise the crystal phase. The amplitude head uses a softplus nonlinearity to enforce non negativity, and both amplitude and phase are multiplied by a fixed real-space support mask $S(x,y)$, in line with standard CXDI support constraints \citep{fienup1982phase,Marchesini2003}. For experimental data, Fourier ViT uses the same support as the MATLAB iterative reconstruction (see Sec.~\ref{sec:Iterative method}) for a fair comparison. Inside the support, the predicted amplitude is blended with a simple prior via a scalar schedule $\alpha(t)$ and then softly rescaled so that its $\ell_2$ norm matches that of the prior amplitude; the precise blending and normalisation rules are given in Supplementary Note 2. For the phase, the two decoder outputs are first normalised onto the unit circle, $(\tilde{c},\tilde{s}) = (\hat{c},\hat{s}) / \sqrt{\hat{c}^{2}+\hat{s}^{2}}$, and the phase is recovered as $\phi(x,y)=\operatorname{atan2}(\tilde{s}(x,y),\tilde{c}(x,y))$.

Given an input diffraction pattern magnitude $M(\mathbf{q})$, the network therefore outputs a complex real-space density $\rho(x,y)$ on a $64\times 64$ grid, with amplitude and phase constrained by the fixed support.

\subsubsection{Loss function}
\label{sec:loss}
For a given input diffraction pattern, the network predicts a complex real-space density $\rho(x,y)$ from the crystal's amplitude and phase. This estimate is propagated through the BCDI forward model to obtain a predicted diffraction magnitude $M_{\mathrm{pred}}(\mathbf{q})$ (Sec.~\ref{sec:decoder}), which is compared with the measured magnitude $M_{\mathrm{meas}}(\mathbf{q})$ via a hybrid Fourier-space loss. The loss combines Pearson correlation coefficient (PCC), an root mean square (RMS)-normalised $\chi^2$, a power weighted $\chi^2$ term and a small total variation (TV) regulariser, with epoch-dependent weights that shift the emphasis from global pattern correlation to fine scale intensity agreement as training proceeds.

For the $\chi^2$-type terms we work with an RMS-normalised magnitude
$\hat{M}(\mathbf{q}) =
\frac{M(\mathbf{q})}{\sqrt{\langle M(\mathbf{q})^{2}\rangle_{\mathbf{q}}}},
$
applied separately to $M_{\mathrm{meas}}$ and $M_{\mathrm{pred}}$ so that any global intensity scale is removed. At training epoch $t$ we minimise,
\begin{equation}
L_{\mathrm{total}}(t) =
w_{\mathrm{PCC}}(t)\,L_{\mathrm{PCC}}
+ w_{\chi^2}(t)\,L_{\chi^2}
+ w_{\mathrm{p}\chi^2}(t)\,L_{\mathrm{p}\chi^2}
+ \lambda_{\mathrm{TV}}(t)\,L_{\mathrm{TV}},
\end{equation}

where the data-fidelity weights \(w_{\mathrm{PCC}}(t)\), \(w_{\chi^2}(t)\), \(w_{\mathrm{p}\chi^2}(t)\) and the TV weight \(\lambda_{\mathrm{TV}}(t)\) follow simple epoch-dependent schedules. They are designed so that PCC dominates early training, power–\(\chi^2\) ramps up over the early epochs to enhance low frequencies, and standard \(\chi^2\) is slightly favoured in the late refinement stage, while TV remains a small stabilising term. The full schedule definitions are given Supplementary Note 3.

\noindent\textbf{(i) PCC loss.}
To enforce global similarity of the diffraction patterns we use
\begin{equation}
L_{\mathrm{PCC}} = 1 - \rho\bigl(M_{\mathrm{meas}}, M_{\mathrm{pred}}\bigr),
\end{equation}
with, for a single pattern,
\begin{equation}
\rho\bigl(M_{\mathrm{meas}}, M_{\mathrm{pred}}\bigr)
=
\frac{\displaystyle\sum_{\mathbf{q}}
       \bigl(M_{\mathrm{meas}}(\mathbf{q}) - \mu_{\mathrm{meas}}\bigr)
       \bigl(M_{\mathrm{pred}}(\mathbf{q}) - \mu_{\mathrm{pred}}\bigr)}
{\sqrt{\displaystyle\sum_{\mathbf{q}}
       \bigl(M_{\mathrm{meas}}(\mathbf{q}) - \mu_{\mathrm{meas}}\bigr)^{2}}\,
 \sqrt{\displaystyle\sum_{\mathbf{q}}
       \bigl(M_{\mathrm{pred}}(\mathbf{q}) - \mu_{\mathrm{pred}}\bigr)^{2}}},
\end{equation}
where $\mu_{\mathrm{meas}}=\langle M_{\mathrm{meas}}(\mathbf{q})\rangle_{\mathbf{q}}$,
$\mu_{\mathrm{pred}}=\langle M_{\mathrm{pred}}(\mathbf{q})\rangle_{\mathbf{q}}$.

\noindent\textbf{(ii) RMS-normalised $\chi^2$ loss.}
To penalise absolute mismatches while remaining insensitive to any residual scale factor, we define
\begin{equation}
L_{\chi^2} =
\frac{\displaystyle\sum_{\mathbf{q}}
       \bigl[\hat{M}_{\mathrm{meas}}(\mathbf{q}) - \hat{M}_{\mathrm{pred}}(\mathbf{q})\bigr]^{2}}
     {\displaystyle\sum_{\mathbf{q}}\hat{M}_{\mathrm{meas}}(\mathbf{q})^{2}},
\end{equation}
where $\hat{M}_{\mathrm{meas}}$ and $\hat{M}_{\mathrm{pred}}$ are the RMS-normalised magnitudes defined above.

\noindent\textbf{(iii) Power-$\chi^2$ loss.}
To give additional emphasis to bright and high-$q$ fringes, we include a power-weighted $\chi^2$ term. Starting from the RMS-normalised magnitudes, we form powered magnitude of measured and predicted diffraction pattern, $\hat{M}_{\mathrm{meas}}^{(p)}(\mathbf{q}) =
\bigl[\hat{M}_{\mathrm{meas}}(\mathbf{q})\bigr]^{p(t)}$ and
$\hat{M}_{\mathrm{pred}}^{(p)}(\mathbf{q}) =
\bigl[\hat{M}_{\mathrm{pred}}(\mathbf{q})\bigr]^{p(t)}$, then define
\begin{equation}
L_{\mathrm{p}\chi^2} =
\frac{\displaystyle\sum_{\mathbf{q}}
       \bigl[\hat{M}_{\mathrm{meas}}^{(p)}(\mathbf{q}) -
             \hat{M}_{\mathrm{pred}}^{(p)}(\mathbf{q})\bigr]^{2}}
     {\displaystyle\sum_{\mathbf{q}}
       \bigl[\hat{M}_{\mathrm{meas}}^{(p)}(\mathbf{q})\bigr]^{2}}.
\end{equation}
The exponent $p(t)$ decays smoothly from about $2$ at early epochs towards $0.5$ later in training, so that optimisation initially focuses on fitting the brightest regions and progressively places more weight on weaker fringes.

\noindent\textbf{(iv) Total-variation amplitude regularisation.} Let \(S(x,y)\in\{0,1\}\) be the binary support mask and \(N_S = \sum_{x,y} S(x,y)\) the number of support pixels. We define an anisotropic TV penalty 
\begin{equation}
L_{\mathrm{TV}} = \frac{1}{N_S} \sum_{x,y} S(x,y)\,\bigl(|\nabla_x A(x,y)| + |\nabla_y A(x,y)|\bigr), 
\end{equation}
where \(\nabla_x A(x,y) = A(x+1,y)-A(x,y)\) and \(\nabla_y A(x,y) = A(x,y+1)-A(x,y)\). In implementation, we mask the amplitude by \(S(x,y)\) before taking finite differences, so the gradients across the support boundary do not contribute.

\begin{figure*}[ht]
    \centering
    \includegraphics[width=12cm]{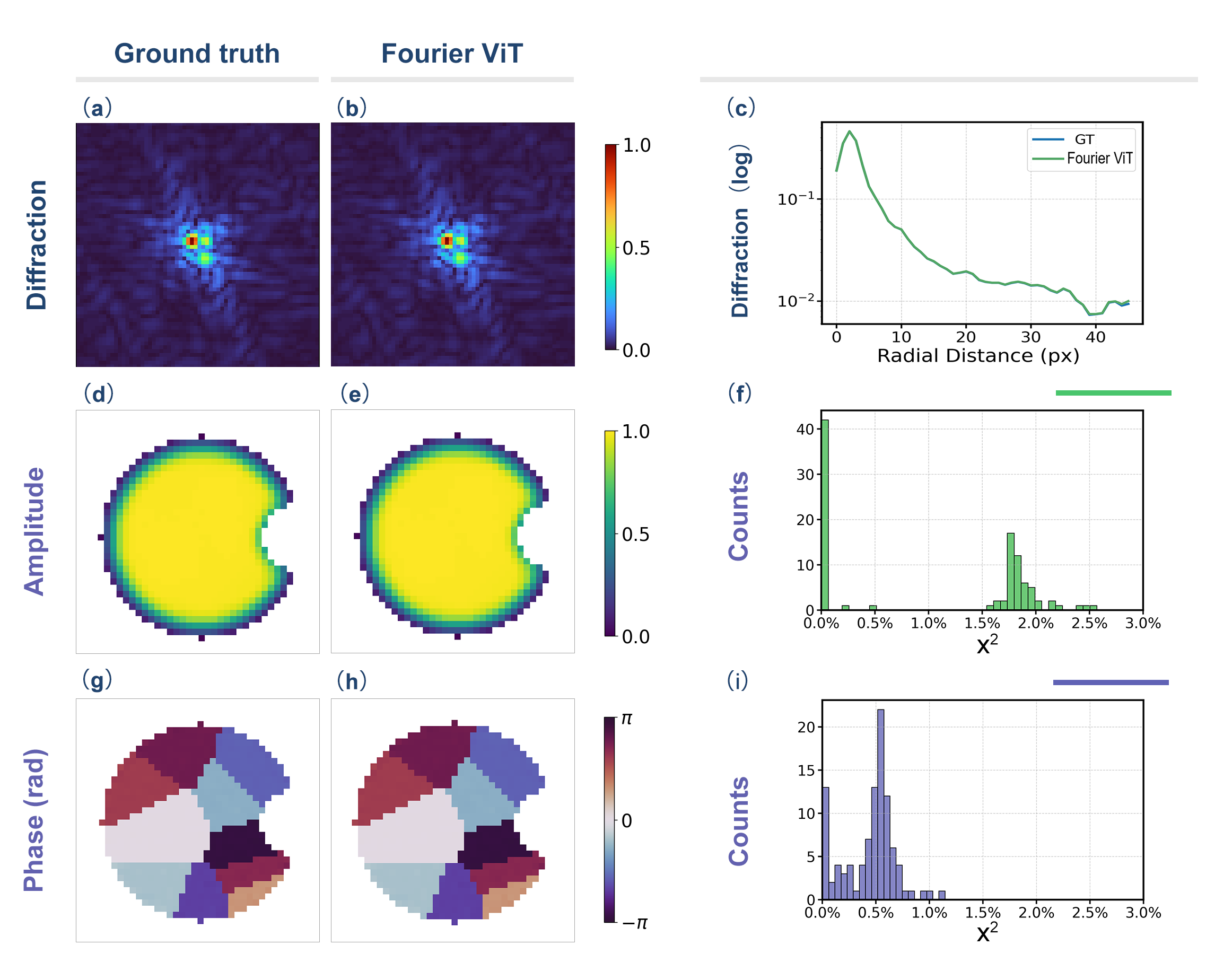}
    \caption{Fourier ViT reconstruction of a synthetic 10-domain crystal from diffraction only. (a,d,g) Ground truth diffraction magnitude (64×64 pixels), crystal amplitude and phase (40×40 pixels); (b,e,h) corresponding Fourier ViT reconstructions. (c) Radial log intensity against $q$ profiles of ground truth and reconstructed diffraction, showing agreement across the full $q$-range. (f,i) Histograms of diffraction $\chi^{2}$ over 100 random initialisations for known crystal amplitude (f) and unknown amplitude (i) phase retrieval.}
    \label{fig:final_3x3}
\end{figure*}

\subsection{Performance on synthetic data}
On synthetic multi-domain crystals, the Fourier ViT recovers the real-space object from reciprocal-space diffraction magnitude, without real-space supervision or support updates. At $64\times64$ diffraction resolution with known crystal amplitude, the method resolves up to 19 domains with runs that achieve perfect convergence, here defined as $\chi^{2}\!\le\!10^{-5}$. Perfect reconstructions occur 42 times out of 100 runs for 10 domains, 18 times for 15 domains, and 4 times for 19 domains. When crystal amplitude and phase are recovered jointly, convergence is harder but still feasible: for a 10-domain crystal, 12 out of 100 runs reach $\chi^{2}\!\le\!10^{-5}$ after 1000 epochs.

To illustrate object amplitude-phase retrieval, we show a representative 10-domain case in Fig.~\ref{fig:final_3x3}. Figs~\ref{fig:final_3x3} (a,b) compare the ground truth and Fourier ViT predicted diffraction magnitudes, while Fig.~\ref{fig:final_3x3}(c) shows the radially averaged log-intensity profile of the diffraction pattern as a function of momentum transfer \(q\) (radial coordinate in reciprocal-space). The profile matches the ground truth across the full \(q\) range, including the high-\(q\) tails produced by strong-phase contrasts and sharp domain walls.
Instead of a single central peak, familiar in the weak-phase case, this pattern has multiple centres because the object has highly varying phase values. Figs~\ref{fig:final_3x3} (d,e) demonstrate that the recovered crystal amplitude is smoothly matching the ground truth, and Figs.~\ref{fig:final_3x3} (g,h) show that the phase solution recovers the real-space domain configurations with sharp, correctly placed boundaries and no spurious domains. The ground truth panels are used only for quantitative assessment and visualisation, as the network is trained unsupervised from diffraction data without real-space labels. For visual comparison, we fix the global phase offset by setting the phase at the central pixel to zero in all reconstructions.

Fig.~\ref{fig:final_3x3}(f) shows a bimodal distribution over 100 random initialisations in the phase-only prediction case. Many runs converge to global minimum solutions with $\chi^{2}\!\le\!10^{-5}$, while the remainder settle into higher-error local minima to form a second cluster at $\chi^{2}\approx (1.5$--$2.5)\times 10^{-2}$. This reflects the non-convex strong-phase retrieval landscape: small shifts in domain wall placement or topology can produce a structured mismatch in the high-$q$ fringe field. With fixed amplitude, the optimisation has limited freedom to accommodate, so these solutions retain stable nonzero residuals.
In joint amplitude-phase recovery in Fig.~\ref{fig:final_3x3}(i), the extra degrees of freedom from amplitude prediction reduce the number of perfectly converged runs but concentrate the remaining outcomes into a narrower band of lower $\chi^{2}$. Here, partially incorrect phase maps can be offset by amplitude reweighting within the support to better match the diffraction magnitude.

These results show that reciprocal-space modelling with Fourier attention reproduces the diffraction structure and recovers the multi-domain phase topology without real-space labels. Unless otherwise stated, all Fourier ViT models were trained in an unsupervised manner using the optimisation and loss schedule described in Sec.~\ref{sec:nn_training}.

\begin{figure*}[ht]
    \centering
    \includegraphics[width=\textwidth]{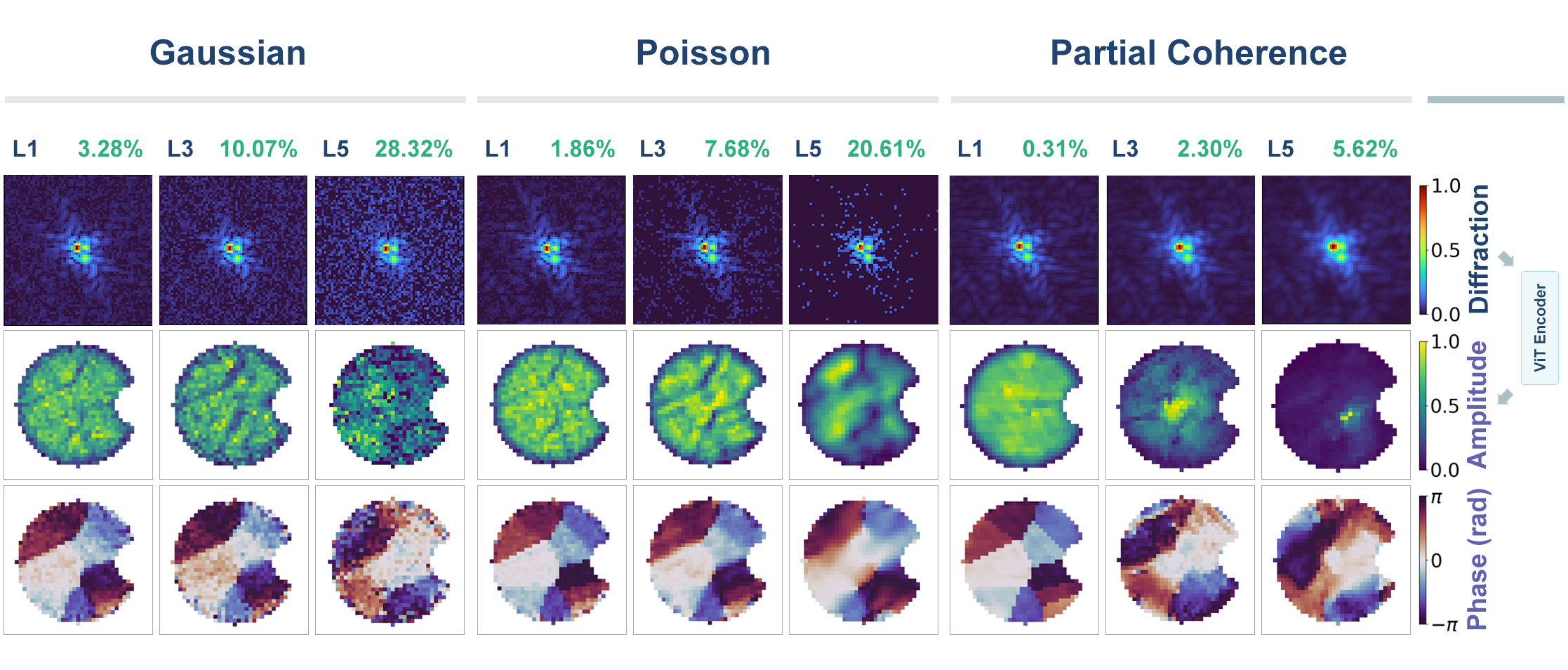}
    \caption{Noise robustness of the Fourier ViT. Columns show three types of noise applied to the same diffraction pattern: Gaussian noise, Poisson noise, and partial coherence. For each model, L1, L3 and L5 denote increasing noise strengths (levels 1, 3 and 5); green percentages give $\chi^2_{\mathrm{n}}$ (in \%), the input diffraction error relative to the clean ground truth. Top row shows the noisy input diffraction amplitudes (64×64 pixels). Middle and bottom rows show the reconstructed crystal amplitude and phase zoomed in 40×40 pixels, respectively; both are masked to the support and phase is in radians.}
    \label{fig:noise_test}
\end{figure*}

For the same simulated diffraction pattern of a 10-domain crystal and fixed support, iterative methods combining Error Reduction (ER) and Hybrid Input-Output (HIO) reach $\chi^2$ values between $10^{-4}$ and $10^{-3}$ for approximately half of reconstructions after 500 iterations. No runs reach our ``perfect'' criterion $\chi^{2}\!\le\!10^{-5}$ with the iterative method, but increasing to 2000 iterations raises the fraction of runs with $10^{-4}\!\le\!\chi^{2}\!\le\!10^{-3}$ to $90\%$. Under matched initialisation and support constraints, the Complex Convolutional Neural Network (C-CNN) with skip connection achieves much higher $\chi^2$ values of $6\times10^{-3}$--$1.2\times10^{-2}$. Among all three methods, the $\chi^{2}$ outcomes over random initialisations are bimodal; however, only the Fourier ViT reaches the lowest-$\chi^{2}$ value ($\chi^{2}\!\le\!10^{-5}$), whereas the C-CNN remains confined to a higher-error mode.

\begin{table}[t]
\centering
\small
\caption{Noise robustness of the Fourier ViT (magnitude domain). All values are max-normalised $\chi^2$ errors in \%. $\chi^2_{\mathrm{n}}$ compares the noisy input to the clean ground truth; $\chi^2_{\mathrm{rec,n}}$ and $\chi^2_{\mathrm{rec,c}}$ compare the reconstruction to the noisy input and to the clean ground truth, respectively. Noise strength increases from level 1 to level 5.}
\renewcommand{\arraystretch}{1.18}
\setlength{\tabcolsep}{3.2pt}

\begin{tabular}{@{}l@{\hspace{20pt}}
                ccc @{\hspace{20pt}}
                ccc @{\hspace{20pt}}
                ccc @{}}
\toprule
& \multicolumn{3}{c}{\textbf{Gaussian}} &
  \multicolumn{3}{c}{\textbf{Poisson}} &
  \multicolumn{3}{c}{\textbf{Partial coherence}} \\
\cmidrule(lr){2-4}\cmidrule(lr){5-7}\cmidrule(lr){8-10}
\textbf{Noise}
& $\chi^2_{\mathrm{n}}$ & $\chi^2_{\mathrm{rec,n}}$ & $\chi^2_{\mathrm{rec,c}}$
& $\chi^2_{\mathrm{n}}$ & $\chi^2_{\mathrm{rec,n}}$ & $\chi^2_{\mathrm{rec,c}}$
& $\chi^2_{\mathrm{n}}$ & $\chi^2_{\mathrm{rec,n}}$ & $\chi^2_{\mathrm{rec,c}}$ \\
\midrule
Level 1 &  3.28 &  1.76 &  1.54 &  1.86 &  1.18 &  0.92 &  0.31 &  0.13 &  0.25 \\
Level 2 &  6.55 &  3.48 &  2.71 &  4.15 &  2.67 &  1.92 &  1.50 &  0.25 &  1.94 \\
Level 3 & 10.07 &  5.16 &  4.14 &  7.68 &  4.69 &  3.65 &  2.30 &  0.23 &  2.78 \\
Level 4 & 18.84 &  9.03 &  8.21 & 12.90 &  8.07 &  5.82 &  3.95 &  0.17 &  4.66 \\
Level 5 & 28.32 & 12.49 & 14.36 & 20.61 & 11.91 & 10.53 &  5.62 &  0.11 &  6.37 \\
\bottomrule
\end{tabular}
\label{tab:noise_mag_levels}
\end{table}

\subsection{Noise robustness}
To assess robustness under realistic experimental degradations, the same noise-free diffraction pattern used in Fig.~\ref{fig:final_3x3} was corrupted by three noise models: additive Gaussian noise, accounting for detector background; Poisson counting noise representing counting statistics; and partial coherence of the incident beam as demonstrated in Fig.~\ref{fig:noise_test}. Strictly, partial coherence is not the same as the statistical noise models because it can be corrected \citep{Clark2012NatCommun}, but its effects on convergence are worth identifying so they can be used for identification. Each noise model was applied at five strengths, from Level 1 to Level 5, according to its effect on $\chi^2$. For visual clarity, Fig.~\ref{fig:noise_test} shows Levels 1, 3 and 5; Levels 2 and 4 are provided in Supplementary Fig. 1.

For each noise type, Table~\ref{tab:noise_mag_levels} reports three normalised diffraction errors: (i) the noisy input relative to the clean ground truth, $\chi^2_{\mathrm{n}}$; (ii) the reconstruction mismatch to the noisy input, $\chi^2_{\mathrm{rec,n}}$; and (iii) the reconstruction error relative to the clean ground truth, $\chi^2_{\mathrm{rec,c}}$. In our unsupervised setting, we select the best run across random initialisations by minimising $\chi^2_{\mathrm{rec,n}}$, and report the corresponding $\chi^2_{\mathrm{rec,c}}$ as evaluation against the clean ground truth.

For Gaussian noise, $\chi^2_{\mathrm{n}}$ increases from 3.28\% to 28.32\% with noise strength. The Fourier ViT reduces the error relative to the clean diffraction to $\chi^2_{\mathrm{rec,c}}=1.54$--14.36\%, corresponding to a 49--59\% reduction relative to $\chi^2_{\mathrm{n}}$. For Poisson noise, $\chi^2_{\mathrm{n}}$ increases from 1.86\% to 20.61\%, and the model reduces this to $\chi^2_{\mathrm{rec,c}}=0.92$--10.53\%, giving a comparable 49--55\% reduction. Across both stochastic noise processes, the reconstruction remains closer to the clean diffraction than the corrupted input itself, demonstrating genuine denoising rather than simply reproducing the noise statistics.

Fig.~\ref{fig:noise_test} illustrates the effects of Gaussian and Poisson noise on the real-space images. At low noise levels, the fringes of the diffraction remain sharp and the reconstructions show clean and domain-resolved crystal phase maps. As noise increases, pixel-scale fluctuations appear and progressively corrupt the phase, eventually blurring or erasing the domain boundaries.
Under Gaussian noise, pixel-scale intensity fluctuations developed in both the reconstructed crystal's amplitude and phase, while under Poisson statistics the amplitude develops grain boundary contrast that mirrors the underlying domain structure. In both cases, the Fourier ViT suppresses unstructured speckle and background fluctuations while preserving the physically meaningful fringe pattern around the centre of the Bragg peak, thereby enabling robust phase retrieval even under strongly degraded conditions. Consistent with Table~\ref{tab:noise_mag_levels}, the Fourier ViT therefore acts like a filter, reducing the diffraction domain error by roughly a factor of two in $\chi^2$ relative to the noisy input.

For the partial coherence condition, $\chi^2_{\mathrm{n}}$ increases from 0.31\% to 5.62\% with increasing blur. The reconstruction matches the blurred measurement extremely closely as $\chi^2_{\mathrm{rec,n}}=0.11$--0.25\%, but the error relative to the clean diffraction lies in the range $\chi^2_{\mathrm{rec,c}}=0.25$--6.37\%. The ratio $\chi^2_{\mathrm{rec,c}}/\chi^2_{\mathrm{n}}$ is $\approx 0.81$ at the weakest level, but increases to $\approx 1.13$--1.29 at stronger blur. The diffraction fringes are progressively blurred and power is concentrated near the central Bragg peak. The reconstructed amplitude shows an over-localised central intensity (“hot spot” region) and a smoothed phase, reflecting the loss of high-$q$ information. Partial coherence is therefore a natural physical origin for the central hot spot in the reconstructed crystal amplitude, a behaviour documented by \citep{Vartanyants2001Partial} and also observed in the experimental reconstructions discussed in Sec.~\ref{sec:experimental_verification}.

\begin{figure*}[ht]
    \centering
    \includegraphics[width=\textwidth]{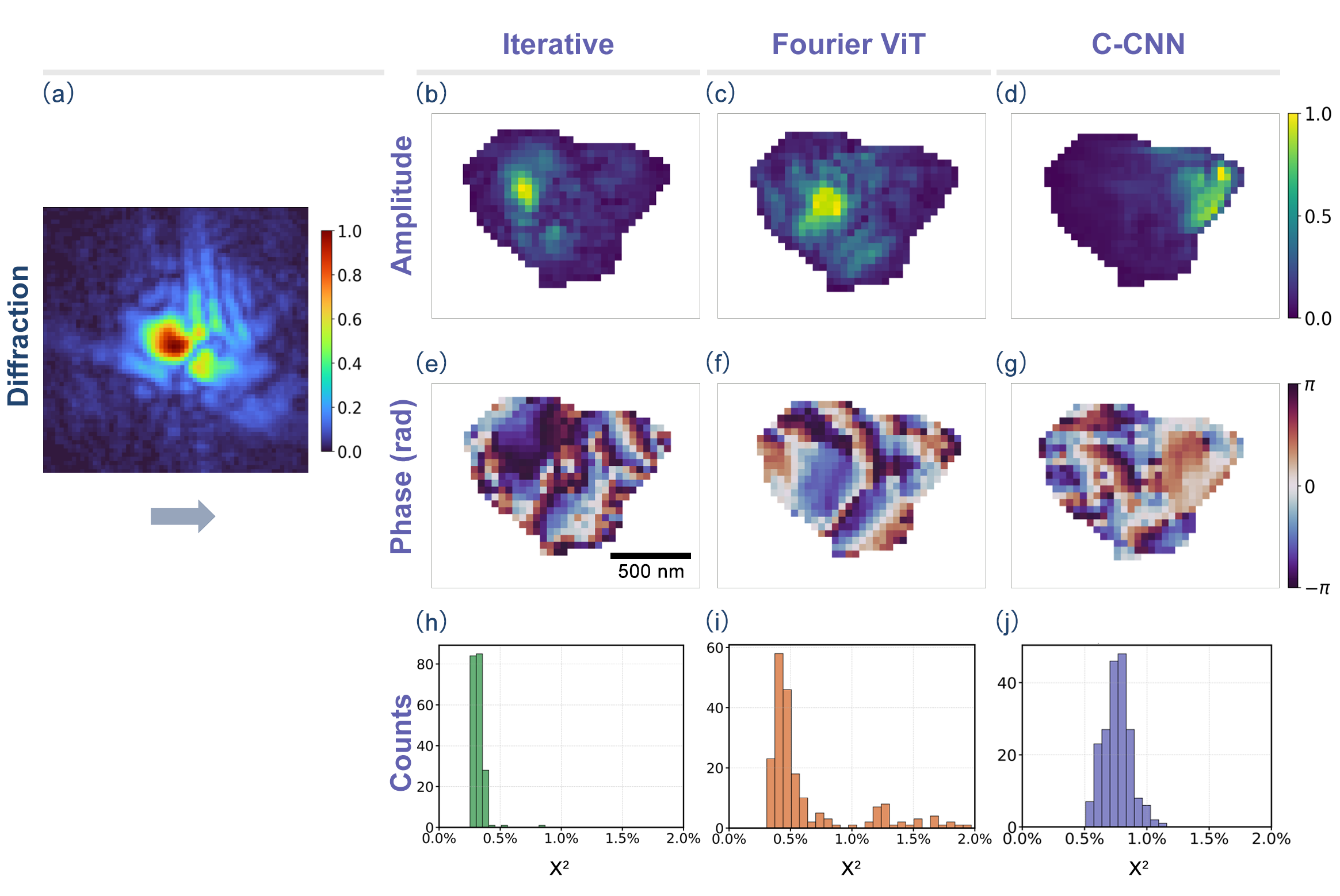}
    \caption{Comparison study of phase retrieval on experimental data from a multi-domain LCMO-500 nanocrystal. (a) Measured BCDI diffraction magnitude (64×64 pixels). (b--d) Reconstructed crystal's amplitude and (e--g) phase from the iterative method, Fourier ViT, and C-CNN. Scale bar is 500\,nm. (h--j) Corresponding $\chi^{2}$ histograms over 200 runs per method, each run with a random initial phase. $\chi^{2}$ range is between $0.0\%$ and $2.0\%$.}

    \label{fig:LCMO500}
\end{figure*}

\subsection{Experimental verification}
\label{sec:experimental_verification}
As a preliminary experimental check, we tested the Fourier ViT on a 2D diffraction pattern from a weak-phase, single-domain $450$ nm \(\mathrm{SrTiO_3}\) (STO) nanocrystal; results are shown in Supplementary Fig. 2. In Fig.~\ref{fig:sto}(a,b), the reconstructions recover a single-domain morphology across runs. Across 100 random initialisations, the final reciprocal-space mismatch \(\chi^{2}\) falls within a narrow band (\(\sim 0.35\%\)–\(\sim 0.68\%\); Fig.~\ref{fig:sto}(c)), indicating reliable convergence in the weak-phase regime. This contrasts with the strong-phase, multi-domain case discussed next, where multiple local minima lead to a substantially broader spread of \(\chi^{2}\) values across runs. 

We then consider the strong-phase regime using an experimental BCDI pattern from a strongly distorted, multi-domain LCMO-500 nanocrystal (Sec.~\ref{sec:bcdi_experiment}). Fig.~\ref{fig:LCMO500} compares reconstructions from a traditional iterative method (ER/HIO), the Fourier ViT, and a C-CNN baseline, using an identical real-space support. For each method, we report the best run (lowest \(\chi^{2}\)) over 200 random initialisations.

For the raw measured diffraction magnitude in Fig.~\ref{fig:LCMO500}(a), the iterative reconstruction (Fig.~\ref{fig:LCMO500}(b,e)) recovers the multi-domain phase structure, but shows isolated hot spots within the amplitude map. The best result reaches $\chi^{2}=0.25\%$ with $\mathrm{PCC}=99.77\%$ against the measured magnitude. Trained unsupervised on the same single diffraction pattern and constrained by the same support, the Fourier ViT returns the amplitude and phase in Fig.~\ref{fig:LCMO500}(c,f). Compared with (b), the Fourier ViT amplitude map (c) shows a broader, less compact high-amplitude region, with the overall amplitude extending over a larger fraction of the crystal; compared with (d), it is less edge-localised, and is more likely to reflect a 2D projection of the object. In the phase map, the recovered domain pattern (f) is consistent with the iterative solution (e), with clearer domain boundaries, more spatially coherent phase regions, and fewer isolated high-contrast features.
The best Fourier ViT run attains $\chi^{2}=0.30\%$ and $\mathrm{PCC}=99.79\%$. By contrast, the best C-CNN result in Fig.~\ref{fig:LCMO500}(d,g) satisfies the support constraint but converges to a different solution: the amplitude concentrates near the support boundary and the phase topology deviates from both the iterative and Fourier ViT reconstructions. This behaviour is accompanied by a higher $\chi^{2}=0.50\%$, suggesting a poorer match to the measured diffraction magnitude under the same forward model.

To remove the hot spot in the crystal amplitude, we tested a mild partial-coherence correction by Gaussian deconvolution of the measured diffraction magnitude \citep{Vartanyants2001Partial}. While the deconvolution visibly sharpens the fringes, it also amplifies other detector noises, and the reconstruction $\chi^{2}$ increases relative to the raw-data case: iterative gives $\chi^{2}=0.61\%$, Fourier ViT has $\chi^{2}=0.57\%$, without significantly improving the hot spot. We therefore use the raw measured pattern as the primary basis for comparison in Fig.~\ref{fig:LCMO500}.

The $\chi^{2}$ histograms in Fig.~\ref{fig:LCMO500}(h--j), plotted on a common axis, summarise run-to-run variability over 200 independent runs per method. Iterative reconstructions (h) cluster tightly within $\chi^{2}\approx 0.25$--$0.87\%$, indicating stable convergence. The Fourier ViT (i) achieves comparable best-case values $\chi^{2}=0.3\%$ but shows a broader spread ($\sim 0.3$--$2.0\%$), while the C-CNN distribution (j) is shifted to higher errors ($\sim 0.5$--$1.1\%$). Supplementary Movies visualise the per-epoch evolution for the three methods and highlight distinct optimisation trajectories.

Overall, on this experimental dataset, the Fourier ViT matches the measured diffraction at least as well as the best iterative reconstruction and outperforms the C-CNN baseline in $\chi^{2}$, while producing multi-domain crystal's amplitude and phase that remain consistent with the iterative solution under identical support constraints.

\section{Discussion}\label{sec:discussion}

We developed a Vision Transformer with Fourier attention (Fourier ViT) for multi-domain phase retrieval in BCDI. 
On synthetic $64\times64$ diffraction patterns, the model recovers strong-phase domain structure directly from diffraction in an unsupervised setting. 
For phase-only prediction with known amplitude, runs can reach essentially perfect agreement with the target diffraction ($\chi^{2}\!\le\!10^{-5}$) while resolving complex domain structure up to 19 domains (Fig.~\ref{fig:final_3x3}f). 
For joint amplitude-phase prediction, a representative 10-domain case reproduces the ground-truth diffraction magnitude over the full $q$ range, including the high-$q$ tails (Fig.~\ref{fig:final_3x3}c), and the reconstructed crystal phase recovers the same domain layout with sharp domain walls (Fig.~\ref{fig:final_3x3}h).
Under the same fixed support on the same pattern, ER/HIO reaches $\chi^{2}=10^{-4}$--$10^{-3}$ for approximately half of reconstructions after 500 iterations, and with sufficient iterations a larger fraction of runs fall into this range.
The complex convolutional neural network (C-CNN) remains at higher $\chi^{2}$ error ($6\times10^{-3}$--$1.2\times10^{-2}$) under the same initial setup and does not reach the $\chi^{2}\!\le\!10^{-5}$ regime achieved by Fourier ViT with sufficient epochs. 
Across all three methods, random initialisation in the strong-phase regime separates runs into low- and higher-$\chi^{2}$ outcomes under the same constraints.

Robustness tests in Fig.~\ref{fig:noise_test} and Table~\ref{tab:noise_mag_levels} show that, under Gaussian and Poisson noise, the reconstructions remain closer to the clean diffraction than the noisy input across all levels ($\chi^2_{\mathrm{rec,c}}<\chi^2_{\mathrm{n}}$), with $\chi^{2}_{\mathrm{rec,c}}$ reduced by roughly a factor of two relative to $\chi^{2}_{\mathrm{n}}$.
This indicates denoising in the diffraction magnitude. By contrast, partial coherence blurs the diffraction and reduces fringe visibility. Although the model can fit the blurred measurement extremely closely, the reconstructed diffraction departs from the clean target as blur increases.
Partial coherence can also produce an over-localised central amplitude “hot spot”, a behaviour documented previously \citep{Vartanyants2001Partial} and observed in our experimental reconstructions (Sec.~\ref{sec:experimental_verification}). Therefore, an explicit coherence model is needed when the blur is significant. All three distortions appear differently in the predicted images: Gaussian noise introduces noisy local features, Poisson noise produces grainy fluctuations, and partial coherence blurs the fringes and produces less well-defined domain boundaries.

We then evaluated the method on measured diffraction from the LCMO-500 nanocrystal against ER/HIO and the C-CNN, using the same support constraint and the same FFT-based forward model in the loss. The three approaches show distinct optimisation behaviour as demonstrated in Supplementary Movies. In ER/HIO, the negative feedback step helps the iterations escape local minima and reach low-$\chi^{2}$ solutions. The C-CNN tends to build the phase map from smooth initial fields toward sharper domain boundaries, adding spatial detail gradually during training; in our tests, it more often converges to solutions with edge-localised amplitude and a phase topology that differs from the other methods. By contrast, the Fourier ViT couples the diffraction-pattern features globally through token mixing and produces multi-domain configurations early in training. This reduces time spent in over-smoothed intermediate states and moves the optimisation quickly into the relevant strong-phase solution class. 

In Fig.~\ref{fig:LCMO500}, the Fourier ViT reaches $\chi^{2}$ values comparable to the best iterative reconstruction (iterative: $\chi^{2}\approx0.25\%$, Fourier ViT: $\chi^{2}\approx0.30\%$) and improves over the C-CNN baseline ($\chi^{2}\approx0.50\%$). Under the same support constraint, the recovered domain morphology is consistent with the ER/HIO result. Compared with ER/HIO, the Fourier ViT amplitude is less concentrated into a compact hot spot, and the phase shows clearer and more spatially coherent domain regions. The reconstruction error $\chi^{2}$ histograms in Fig.~\ref{fig:LCMO500} (h-j) also highlight a practical aspect of the strong-phase regime. With realistic noise and unavoidable model mismatch under a magnitude-only data constraint, many near-degenerate minima can fit the measured diffraction almost equally well. As a result, reconstructions may be nearly indistinguishable in reciprocal-space while remaining meaningfully different in real-space. Therefore, reducing $\chi^{2}$ is necessary but not sufficient to identify a unique domain topology. The broader spread of Fourier ViT outcomes is more naturally interpreted as rapid access to multiple acceptable strong-phase solutions enabled by global token mixing, rather than as optimisation instability. Practically, we report the best-$\chi^{2}$ reconstructions and verify that key domain-level features are reproducible across independent runs when the measured intensity and imposed constraints do not enforce a unique solution.

Computationally, ER/HIO is dominated by repeated FFT projections, typically one forward and one inverse FFT per iteration \citep{fienup1982phase,Duhamel1990}. 
In the unsupervised learning condition, the FFT forward model is embedded in the loss alongside network forward/backward propagation for both the C-CNN and Fourier ViT. 
Within the Fourier ViT, Fourier token mixing enables global coupling while avoiding the quadratic scaling of dot-product self-attention \citep{Vaswani2017Attention,LeeThorp2022FNet}.

Overall, Fourier ViT offers a practical route to strong-phase, multi-domain BCDI reconstructions. In near-ideal conditions, all methods improve with additional iterations or epochs, but for a fixed optimisation budget the Fourier ViT reaches structured multi-domain solutions earlier and outperforms the C-CNN baseline, while remaining comparable to the best ER/HIO reconstructions on experimental data. Future work will incorporate explicit noise and partial-coherence models in the forward operator, develop uncertainty-aware selection from solution ensembles, learn how to derive the support from the diffaction data and extend the approach to fully 3D experimental datasets.

\begin{figure*}[ht]
    \centering
    \includegraphics[width=10cm]{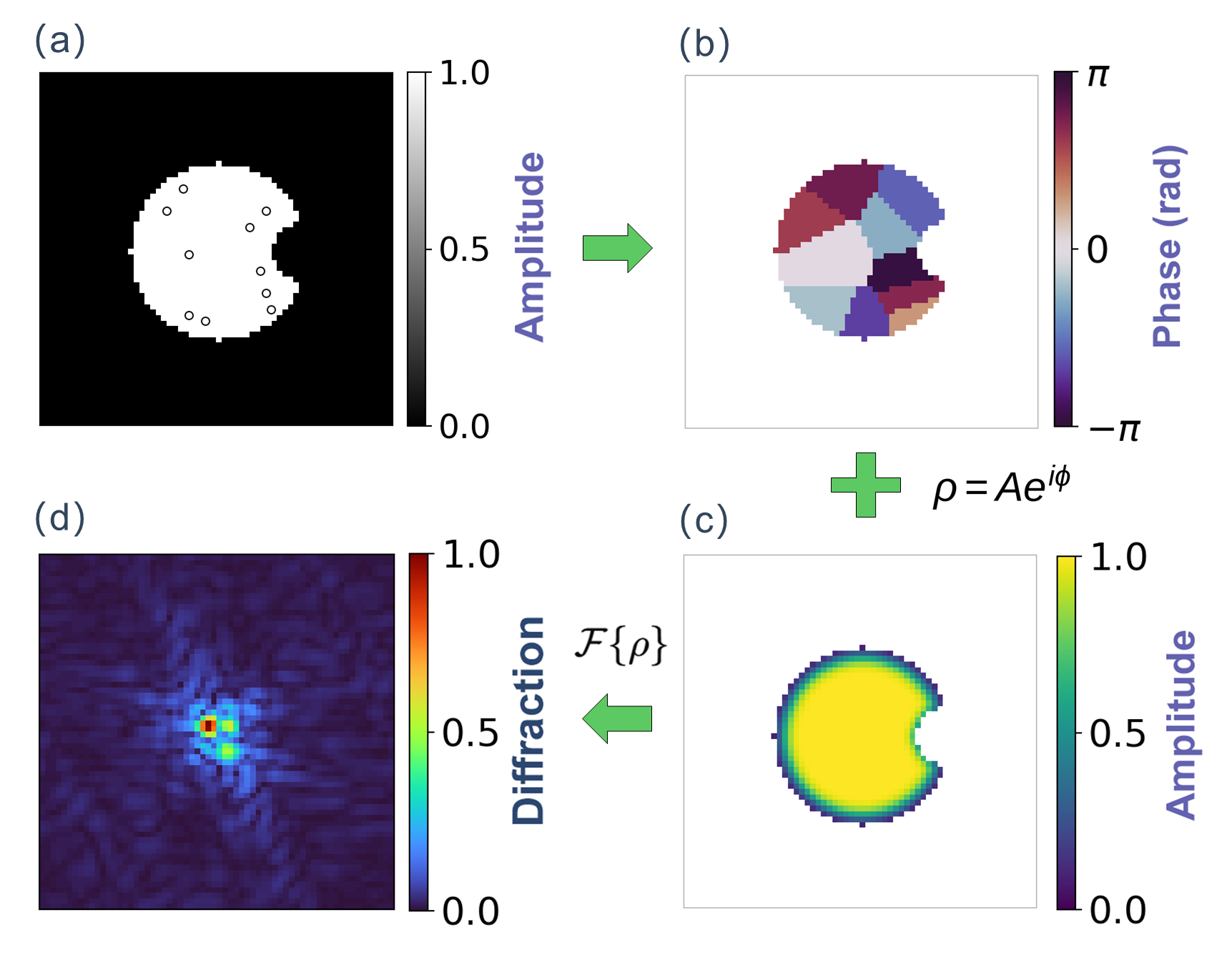}
    \caption{Simulation pipeline for synthetic crystals and diffraction patterns.
    (a) Random seed placement within the support.
    (b) Voronoi phase domain generation from seeds, with phases wrapped to $[-\pi,\pi]$.
    (c) Construction of the crystal amplitude mask, max-normalised to $[0,1]$.
    (d) Corresponding diffraction magnitude computed by fast Fourier transform (FFT), max-normalised to $[0,1]$.
    All images are $64\times64$ pixels.}
    \label{fig:synthetic_data}
\end{figure*}

\section{Methods}\label{sec:methods}
\subsection{Synthetic Data and Noise}\label{sec:synthetic}
\subsubsection{Voronoi multi-domain crystals}\label{sec:voronoi}
To simulate the heterogeneous structure of functional nanocrystals, we constructed two-dimensional crystals made up of multiple phase domains with sharp, straight boundaries. In Fig.~\ref{fig:synthetic_data}(a), seed points were randomly placed inside a soft-edged circular support with a crescent cutout, representing the particle shape, and each seed was assigned a random phase in $[-\pi,\pi]$. The soft amplitude edges suppress artificial high-frequency components in Fourier space that would arise from box-shaped boundaries, while the crescent cut breaks symmetry and helps to avoid twin solutions. A minimum spacing of 2~pixels was enforced between seeds to prevent clustering. Each seed then generated a Voronoi cell whose pixels take the phase of the closest seed, producing a tessellation of non-overlapping domains with straight boundaries. This Voronoi-based construction mimics the abrupt domain walls typically observed in ferroelectric and ferroelastic materials \citep{Setter2006ThinFilms}.

In Fig.~\ref{fig:synthetic_data}(b), we set the central domain to $\phi=0$ to remove the global phase offset. We required $\Delta\phi>0.5$~rad between neighbouring domains to enforce strong-phase contrast and avoid inversion-symmetric cases. Fig.~\ref{fig:synthetic_data}(c) shows the crystal amplitude $A(x,y)$, chosen to represent the 2D projected amplitude of a finite 3D particle. We use a flat-core radial profile with Gaussian edge smoothing, modulated by the crescent mask, to produce a confined but soft-edged support. The final real-space object is defined as the complex density
\begin{equation}
\rho(x,y) = A(x,y)\, e^{i\phi(x,y)},
\end{equation}
with $A(x,y)$ fixed and $\phi(x,y)$ set by the Voronoi partition.

In Fig.~\ref{fig:synthetic_data}(d), the diffraction intensity is obtained as the modulus squared of the Fourier transform of the complex density,
\begin{equation}
I_{\mathrm{pred}}(\mathbf{q}) = \bigl|\mathcal{F}\{\rho(x,y)\}(\mathbf{q})\bigr|^2,
\end{equation}
on a $64\times64$ grid containing a 32-pixel-diameter support, corresponding to an oversampling ratio of 2 in reciprocal-space. Outside the support, $\rho(x,y)$ is set to zero so that only physically allowed regions contribute to the scattering. The crystal amplitude, phase and diffraction pattern are stored as 16-bit TIFF images. Overall, this pipeline produces synthetic data with a controlled number of domains, strong-phase boundaries and realistic reciprocal-space structure, which we use to benchmark Fourier ViT phase retrieval networks.

\subsubsection{Noise tests}\label{sec:noise_tests}

We assess robustness in the regime where the network predicts both crystal amplitude and phase with noise added to the diffraction pattern. Starting from a max-normalised diffraction pattern intensity $I(q_x,q_y)\in [0,1]$, we construct three families of corrupted data: Poisson shot noise, Gaussian read noise and partial coherence.

For the two detector-noise models we work on the exposure (photon-count) scale. In this setting,
read noise from the sensor chain is well approximated by additive, zero-mean Gaussian noise in
counts, whereas photon shot noise is described by Poisson statistics \citep{Goodman2015,Janesick2001,Janesick2007}.
Given a conversion factor \(S>0\) from normalised intensity to expected counts, the ideal noiseless
counts are $C_0(q_x,q_y) = S\,I(q_x,q_y)$.
Gaussian read noise is then modelled as
\begin{equation}
C_{g}(q_x,q_y) = C_0(q_x,q_y) + \eta(q_x,q_y),
\end{equation}
where the random variable \(\eta(q_x,q_y)\) is drawn independently at each pixel from a normal
distribution with zero mean and variance \(\sigma_c^2\). Poisson shot noise is modelled as
\begin{equation}
C_{p}(q_x,q_y) \sim \mathrm{Poisson}\!\bigl(C_0(q_x,q_y)\bigr),
\end{equation}
reflecting the discrete, independent arrival of photons at each pixel. The function of $\mathrm{Poisson}$ is generated using the Poisson random number generator in the NumPy array library \citep{Harris2020ArrayNumPy}. In both cases we enforce the physical non-negativity of the recorded counts by truncating \(C(q_x,q_y)\ge 0\). For analysis, we return to an intensity scale via \(\tilde I(q_x,q_y) = C(q_x,q_y)/S\), and use the corresponding
diffraction magnitude \(\tilde M(q_x,q_y) = \sqrt{\tilde I(q_x,q_y)}\) for all reported image space
metrics and for the reconstruction loss \(\chi^{2}_{\mathrm{rec}}\) in Table~\ref{tab:noise_mag_levels}.

Partial coherence is treated as a degradation of the diffraction intensity rather than a stochastic detector noise. We model it as a contrast-reducing blur
\begin{equation}
I_{\mathrm{pc}}(q_x,q_y) = \mathcal{G}_\sigma * I(q_x,q_y),
\end{equation}
where $\mathcal{G}_\sigma$ approximates the far-field mutual-coherence function as a Gaussian in reciprocal space, so that partial coherence is modelled as a convolution of the coherent intensity \citep{Williams2007PRB,Clark2012NatCommun}. For simulated data, we convolve the coherent diffraction with $\mathcal{G}_\sigma$ and then renormalise $I_{\mathrm{pc}}$ to match the clean pattern’s overall scaling, so that changes in reconstruction quality primarily reflect the loss of high-$q$ fringe visibility rather than intensity rescaling. For experimental data (Sec.~\ref{sec:experimental_verification}), we apply the same Gaussian model in reverse using a regularised deconvolution to partially undo the blurring and obtain a more nearly coherent effective diffraction pattern.

\subsection{Fourier ViT-network training}
\label{sec:nn_training}
Fourier ViT operates on $64\times64$ max-normalised diffraction magnitudes. With a patch size of 4, this gives $16\times16$ tokens and an embedding dimension of 128 processed through 10 Fourier transformer blocks. The networks are trained for 1000 epochs with batch size 1 using AdamW \citep{LoshchilovHutter2019AdamW} (initial learning rate $5\times10^{-4}$, weight decay $10^{-5}$, $\beta_{1}=0.9$, $\beta_{2}=0.999$) with a cosine-annealed learning rate schedule down to $10^{-6}$. Training uses automatic mixed precision on an NVIDIA Quadro RTX 5000 (Max-Q).

The hybrid loss function of Sec.~\ref{sec:loss} is used throughout. A fixed binary support is imposed; the amplitude is initialised from a Gaussian prior within the support and kept fixed for the first 10 epochs, before being learned jointly with the phase via the blending schedule $\alpha(t)$ (Supplementary Note 2). Dropout is set to 0, and we apply gradient clipping with a maximum gradient norm of 1.0 to stabilise optimisation.

For each domain count, we perform 100 independent runs with different random seeds and report statistics over these runs. We select the best run across random initialisations by minimising the diffraction mismatch $\chi^{2}$ between the predicted and measured diffraction magnitudes, and report the corresponding reconstruction metrics, including $\chi^{2}$ against the clean target when available.

\subsection{BCDI experiment}
\label{sec:bcdi_experiment}
We measured BCDI data from a crystal of La$_{0.5}$Ca$_{0.5}$MnO$_3$ (LCMO). LCMO is an important complex transition-metal oxide showing colossal magnetoresistance (CMR) close to room temperature, in which large changes in electrical resistance are induced by small changes in magnetic field. We are interested in whether the phase domains play a role in the CMR behaviour. LCMO was prepared by chemical synthesis in the form of nanocrystals with sizes in the range of 300\,nm, aggregated into clusters \citep{RobinsonPhaseDomains}. These were attached to silicon wafer substrates for measurement by BCDI using the tetraethyl orthosilicate (TEOS) bonding method \citep{Monteforte2016Silica}: a dilute solution of TEOS was used to coat the dry powder, and it was then calcined at 500~\ensuremath{^\circ\mathrm{C}} in air using a furnace for several hours.

X-ray measurements were made at beamline 34-ID-C of the Advanced Photon Source (APS) \citep{RobinsonPhaseDomains}. The monochromatic X-ray beam, with wavelength $\lambda = 0.138\,\mathrm{nm}$, was made coherent with a $50\,\mu\mathrm{m}$ (vertical) $\times$ $20\,\mu\mathrm{m}$ (horizontal) slit, 48\,m from the source, then focused with Kirkpatrick--Baez (KB) mirrors to a beam size of $600\,\mathrm{nm} \times 800\,\mathrm{nm}$ at the centre of a precise air-bearing rotation axis, $\theta$. A single nanocrystal was selected by rotating $\theta$ until an isolated peak appeared on the Medipix area detector, with $55\,\mu\mathrm{m}$ pixels, positioned at the Bragg angle of the $(110)$ Bragg peak, at a distance $D = 2.2\,\mathrm{m}$ away from the sample. A single frame from that 3D data set, LCMO-500, was selected for the study. The corresponding image is a phased projection of the 3D nanocrystal structure, whose Fourier transform magnitude squared (plus noise) should match exactly the measured frame, according to the Fourier-slice theorem.

\subsection{Iterative method}
\label{sec:Iterative method}
We implement a conventional 2D iterative phase retrieval algorithm in MATLAB following standard BCDI practice \citep{RobinsonHarder2009,Clark2012NatCommun}. The initial support is defined relative to the reconstruction grid, with $s_x=0.5\,N_x$ and $s_y=0.5\,N_y$. We then use a predefined switching schedule that alternates between 20 iterations of error-reduction (ER) and 80 iterations of hybrid input-output (HIO) algorithms, with HIO relaxation parameter $\beta=0.9$ \citep{fienup1982phase}. The support was refined using shrink-wrap updates \citep{Marchesini2003} with a max-normalised amplitude threshold $=0.15$. 
During ER, we enforce the crystal's amplitude to zero outside of the support and keep phase within $[-\pi,\pi]$. Reconstructions of LMCO-500 used 300 iterations with shrinkwrap starting at iteration 200 to establish a support, which was then used to compare Fourier ViT, CNN and iterative results. Fixed-support reconstructions of simulated data were performed for up to 2000 iterations to establish convergence statistics.

\subsection{Complex Convolutional Neural Network}
For experimental comparison with the proposed Fourier ViT, we implement a complex-valued CNN baseline following the complex-network recipe used for BCDI phase retrieval \citep{yu2024ultrafast}. The model has a U-Net encoder-decoder architecture. In our implementation, the encoder uses complex channel widths of 32, 64, 128, 256 and 512 (with a decoder of 256, 128, 64 and 32), followed by a final $1\times1$ complex convolution that outputs a single complex channel. We used LeakyReLU activations with a 0.2 negative slope throughout. Feature maps are represented as real and imaginary channels, and all convolutions are performed via complex operations by mixing the two channels according to complex multiplication. Skip connections bridge the encoder and decoder stages to preserve fine spatial detail. For the LCMO-500 measurement, we initialised the network input as two channels, with $\Re\{x\}$ set to the max-normalised measured diffraction magnitude and $\Im\{x\}=0$, corresponding to a zero initial phase. The network outputs a complex object $\rho(x)=\rho_{\Re}(x)+i\rho_{\Im}(x)$, from which amplitude and phase are obtained by $|\rho|$ and $\arg(\rho)$. Training and evaluation follow the same procedure as Sec.~\ref{sec:nn_training}.

\paragraph*{Code availability}
The code used to train and evaluate the Fourier ViT and to reproduce all figures is available at \url{https://github.com/JialunSimonLiu/Fourier-Attention_Vision-Transformer}.

\newpage
\section*{SUPPLEMENTARY NOTE 1: Details of multi-scale Fourier attention}
The spectral mixer introduced in Sec. 2.1.2 is defined here. Let $\mathbf{X}\in\mathbb{R}^{B\times E\times H\times W}$ with $H{=}W{=}16$ denote the token feature map inside a transformer block. At scale $s$ we downsample by average pooling, apply a per-channel Fourier gate, and upsample back:
\[
\mathbf{Z}_s
= \mathcal{U}_s\!\Big(\mathcal{F}^{-1}\!\big(\,\mathcal{F}(\mathcal{D}_s\mathbf{X})\ \odot\ \mathbf{W}_s\ \odot\ \mathbf{M}_s\,\big)\Big),
\]
where $\mathcal{D}_s$ is average pooling with stride $s$ to $h_s{\times}w_s=(H/s){\times}(W/s)$, $\mathcal{U}_s$ upsamples back to $(H,W)$, $\mathcal{F}$ is the channel-wise 2D FFT, $\mathbf{W}_s\in\mathbb{R}^{E\times h_s\times w_s}$ are learned per-channel frequency responses, and $\mathbf{M}_s\in\mathbb{R}^{1\times h_s\times w_s}$ is a shared gate. The multi-scale output is
\[
\mathbf{Z}=\sum_{s}\alpha_s\,\mathbf{Z}_s,\qquad
\boldsymbol{\alpha}=\mathrm{softmax}(\boldsymbol{\theta}),
\]
with scale weights $\alpha_s$ parameterised by $\boldsymbol{\theta}\in\mathbb{R}^{|\{s\}|}$.

By the convolution theorem,
\[
\mathbf{Z}_s=\mathcal{U}_s\big(\,\mathcal{D}_s\mathbf{X}\ \ast\ \mathbf{g}_s\,\big),\quad 
\mathbf{g}_s=\mathcal{F}^{-1}(\mathbf{W}_s\odot\mathbf{M}_s),
\]
with $\ast$ denoting circular convolution on the $h_s{\times}w_s$ grid. Each scale applies the learned kernel $\mathbf{g}_s$, upsamples the result to $(H,W)$, and combines the scales using the weights $\alpha_s$. This gives global token mixing via FFT-based convolution. In frequency space, $\mathbf{W}_s\odot\mathbf{M}_s$ is a multiplicative gate on the Fourier modes.

Let $N=H W$ denote the number of tokens. For each transformer block, the spectral mixer requires $\mathcal{O}(|\{s\}|\,B\,E\,N\log N)$ time and $\mathcal{O}(|\{s\}|\,B\,E\,N)$ memory, compared with $\mathcal{O}(B\,E\,N^2)$ time and $\mathcal{O}(B\,N^2)$ memory for dot-product attention. This follows Fourier-based token mixing (FNet) \citep{LeeThorp2022FNet} and spectral parameterisations used in Fourier neural operators \citep{Li2021FNO}, but in our model it appears only as a compact multi-scale gate inside each transformer block.
\newpage

\subsection*{SUPPLEMENTARY NOTE 2: Amplitude prior and normalisation}
The decoder predicts a real-space amplitude $\hat{A}(x,y)$ on the $64\times 64$ grid, together with a phase field as described in Sec. 2.1.3. Before forming the complex density we combine $\hat{A}(x,y)$ with a simple prior amplitude $A_{\mathrm{prior}}(x,y)$ defined on the real-space support $S(x,y)$, where $S(x,y)=1$ inside the support region and $S(x,y)=0$ outside. From this support we construct one of three priors:
(i) a flat prior with $A_{\mathrm{prior}}(x,y)=S(x,y)$;
(ii) a blurred prior obtained by repeated local averaging of $S(x,y)$; or
(iii) a Gaussian prior centred on the support and truncated by $S(x,y)$.

During training we blend the prior and predicted amplitudes via a scalar weight $\alpha(t)$:
\[
A_{\mathrm{blend}}(x,y;t)
= \alpha(t)\,A_{\mathrm{prior}}(x,y) + \bigl(1-\alpha(t)\bigr)\,\hat{A}(x,y).
\]
This keeps early epochs guided by the support-based prior and later epochs dominated by the network prediction. We use a cosine schedule with a lock epoch $t_{\mathrm{lock}}$ and a decay-end epoch $t_{\mathrm{end}}$:
\[
\alpha(t) =
\begin{cases}
1, & t < t_{\mathrm{lock}},\\[2pt]
\dfrac{1}{2}\Bigl(1+\cos\!\dfrac{\pi\,(t-t_{\mathrm{lock}})}{t_{\mathrm{end}}-t_{\mathrm{lock}}}\Bigr), & t_{\mathrm{lock}} \le t < t_{\mathrm{end}},\\[6pt]
0, & t \ge t_{\mathrm{end}}.
\end{cases}
\]
For the experimental single-pattern runs, we use $t_{\mathrm{lock}}=10$ and $t_{\mathrm{end}}=50$. For the synthetic dataset runs, we use a slower schedule ($t_{\mathrm{lock}}=50$, $t_{\mathrm{end}}=200$) to reduce early amplitude drift in joint amplitude-phase recovery; in all cases the prior is removed once $\alpha(t)=0$.

To stabilise the overall scale of the reconstructed amplitude, we apply a soft energy-matching step inside the support. Let $A_{\mathrm{blend}}(x,y;t)$ denote the blended amplitude at epoch $t$.
We compute the total amplitude “energy” inside the support for the blended and prior amplitudes as
\[
E_{\mathrm{cur}}(t)
= \sqrt{\sum_{x,y}\bigl(A_{\mathrm{blend}}(x,y;t)\,S(x,y)\bigr)^2},\qquad
E_{\mathrm{tgt}}
= \sqrt{\sum_{x,y}\bigl(A_{\mathrm{prior}}(x,y)\,S(x,y)\bigr)^2}.
\]
where $E_{\mathrm{cur}}$ is the current energy, $E_{\mathrm{tgt}}$ is the targeted energy. We then rescale
\[
\tilde{A}(x,y;t) = \gamma(t)\,A_{\mathrm{blend}}(x,y;t),\qquad
\gamma(t) = \mathrm{clip}\bigl(E_{\mathrm{tgt}}/E_{\mathrm{cur}}(t),\,\gamma_{\min},\gamma_{\max}\bigr),
\]
with $\gamma_{\min}{=}0.8$ and $\gamma_{\max}{=}1.2$ in all experiments. This constrains the support energy to remain close to the prior value. In practice, it prevents global amplitude drift while allowing local deviations from $A_{\mathrm{prior}}$.
\newpage

\subsection*{SUPPLEMENTARY NOTE 3: Loss weights and exponents}

The hybrid loss combines Pearson correlation coefficient (PCC), standard \(\chi^2\), and power--\(\chi^2\),
\[
L_{\text{total}}(t)=
w_{\mathrm{PCC}}(t)\,L_{\mathrm{PCC}}
+ w_{\chi^2}(t)\,L_{\chi^2}
+ w_{\mathrm{p}\chi^2}(t)\,L_{\mathrm{p}\chi^2},
\]
with epoch index \(t\ge 0\). We set the raw weight schedules as
\begin{align*}
\text{raw}_{\mathrm{PCC}}(t) &= 200\,\exp(-t/15) + 1, \\
\text{raw}_{\chi^2}(t)      &= 1, \\
\text{raw}_{\mathrm{p}\chi^2}(t) &= 0.5\,r(t),
\end{align*}
where \(r(t)\) is a sine ramp from 0 to 1 over the first 20 epochs. For \(t \ge 200\),
we rescale the raw weights as
\[
\text{raw}_{\mathrm{PCC}}(t) = 0.6\,\text{raw}_{\mathrm{PCC}}(t),\quad
\text{raw}_{\mathrm{p}\chi^2}(t) = 0.6\,\text{raw}_{\mathrm{p}\chi^2}(t),\quad
\text{raw}_{\chi^2}(t) = 1.1\,\text{raw}_{\chi^2}(t).
\]
We then normalise by the sum of raw weights at epoch \(t\) to obtain
\(w_{\mathrm{PCC}}(t)\), \(w_{\chi^2}(t)\), and \(w_{\mathrm{p}\chi^2}(t)\).

The exponent in the power--\(\chi^2\) term decays from 2.0 to 0.5,
\[
p(t)=p_{\mathrm{end}} + \bigl(p_{\mathrm{start}}-p_{\mathrm{end}}\bigr)\exp(-\lambda t),
\]
with \(p_{\mathrm{start}}=2.0\), \(p_{\mathrm{end}}=0.5\), and \(\lambda=0.005\).
We include a total-variation term on the amplitude with weight
\(\lambda_{\mathrm{TV}}(t)=0.01\) for \(t<200\) and \(\lambda_{\mathrm{TV}}(t)=0\) thereafter.
\newpage

\begin{figure*}[ht]
    \centering
    \includegraphics[width=15cm]{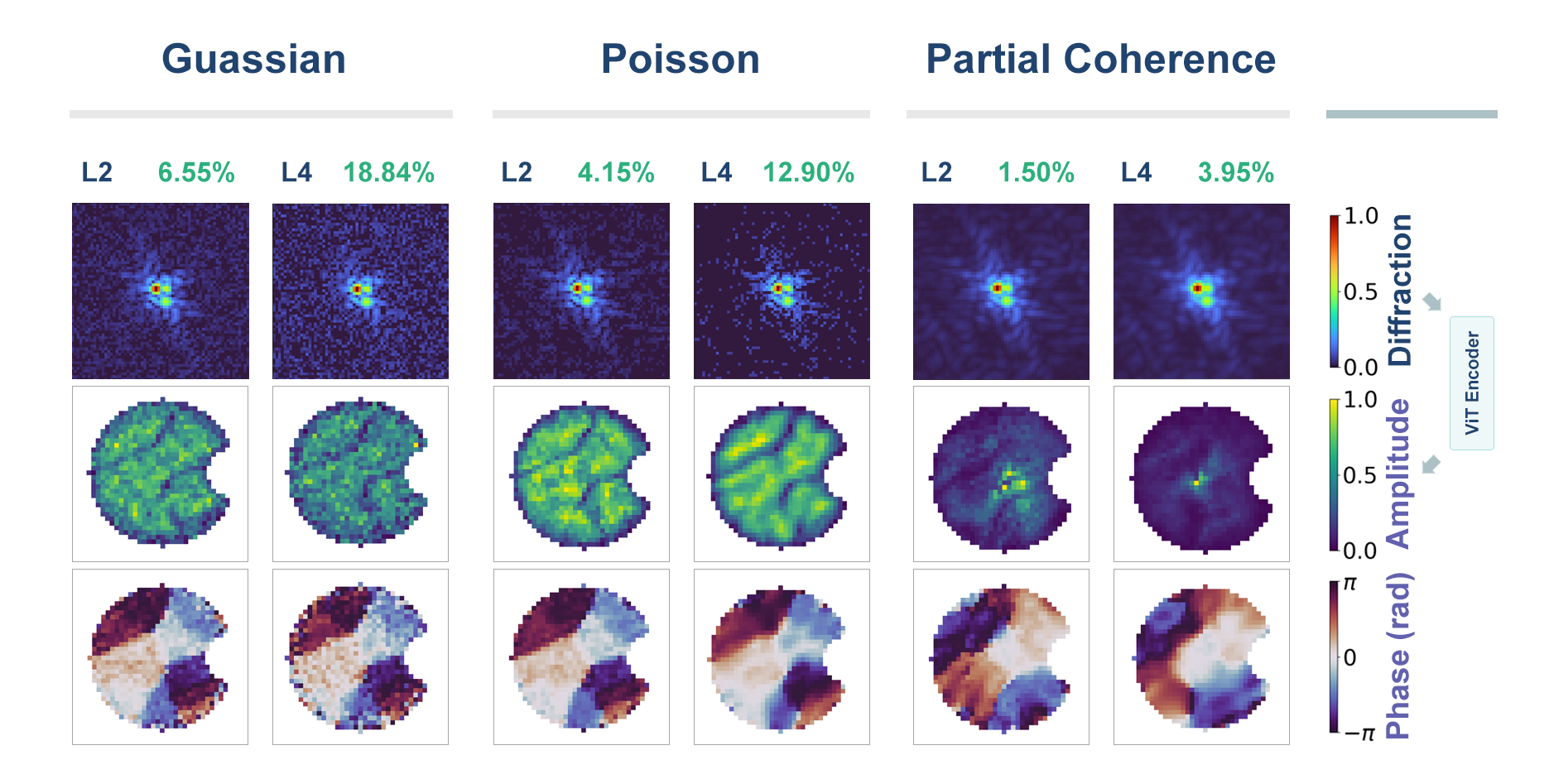}
    \caption{Additional noise robustness visualisations for the Fourier ViT at Levels 2 and 4 (Levels 1, 3 and 5 are shown in Fig. 3). Columns show Gaussian noise, Poisson noise, and partial coherence applied to the same reference diffraction pattern. For each noise model, the green percentages report $\chi^2_{\mathrm{n}}$ (in \%), the input diffraction error relative to the clean ground truth. The top row shows the noisy input diffraction magnitude (64$\times$64 pixels). The middle and bottom rows show the reconstructed crystal amplitude and phase (40$\times$40 pixels), respectively; both are masked to the support and the phase is shown in radians.}
    \label{fig:noise_test_sup}
\end{figure*}

\begin{figure*}[ht]
    \centering
    \includegraphics[width=10cm]{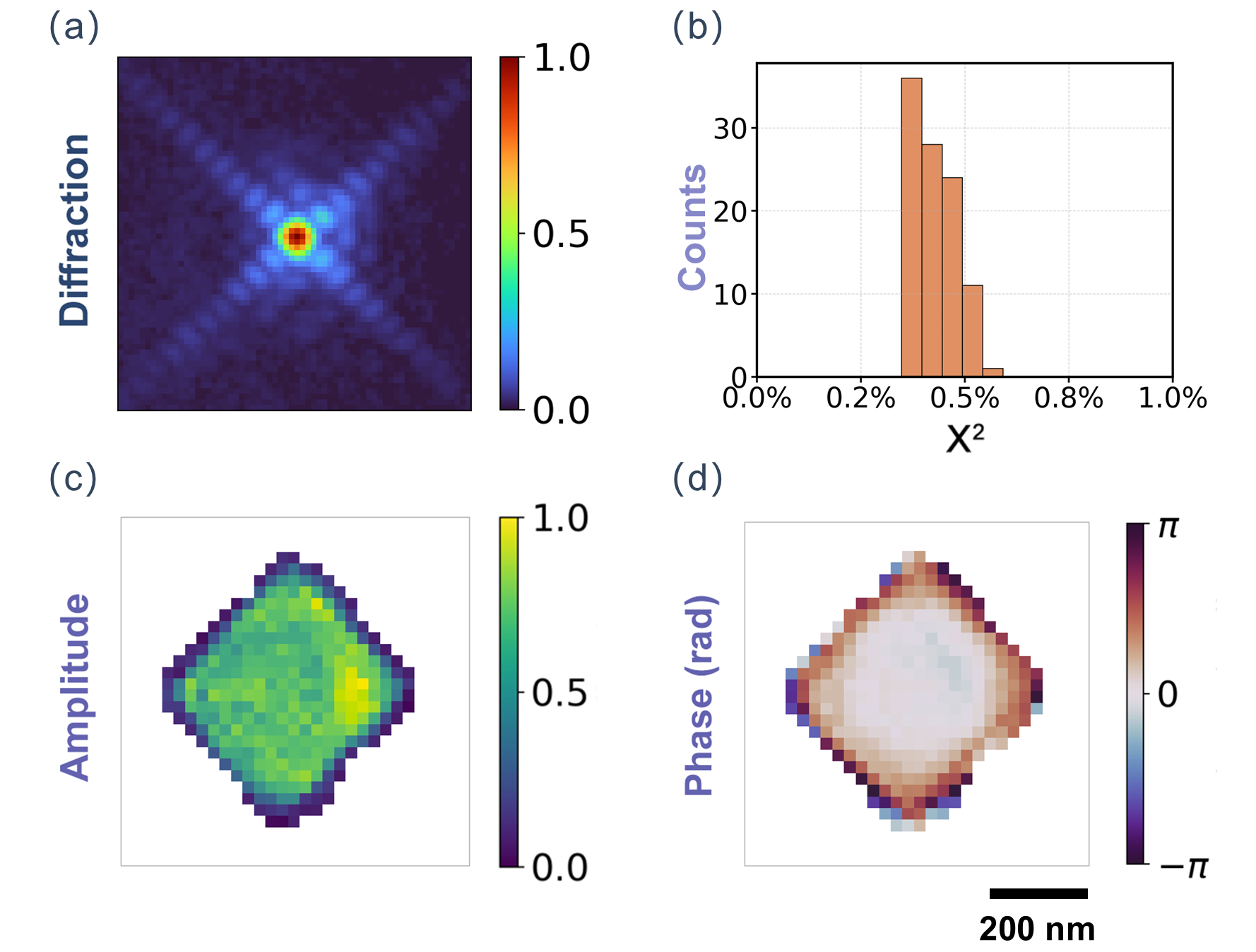}
    \caption{Single-domain SrTiO$_3$ (STO) nanocrystal with a cross-section size of $\sim 460 \times 450$~nm$^2$. (a) Measured STO diffraction magnitude, shown as a $64\times64$ pixel cropped region centred on the Bragg peak and max-normalised to $[0,1]$. (b) Distribution of final reciprocal-space mismatch $\chi^{2}$ over 100 random initialisations, ranging from $\sim 0.35\%$ to $\sim 0.68\%$. (c) Reconstructed real-space amplitude, max-normalised to $[0,1]$. (d) Reconstructed real-space phase, wrapped to $[-\pi,\pi]$; scale bar represents 200 nm.}
    \label{fig:sto}
\end{figure*}
\newpage
\section*{Acknowledgements}
We would like to thank Xi Yu, Wenxuan Fang, Jack Griffiths, Yue Dong, Ruiyang Duan, Seungjun Lee, Yuewei Lin, Longlong Wu and Shinjae Yoo for helpful discussions, and Weibin Peng for advice on figure design and presentation. We would like to thank Ross Harder, Tadesse Assefa, Yue Cao, Xiaojing Huang and Jesse Clark for help with the measurements of LCMO and Josh Turner and Arup Kumar Raychaudhuri for providing the sample. The measurements were carried out at the Advanced Photon Source (APS) beamline 34-ID-C, which was supported by the U.S. Department of Energy, Office of Science, Office of Basic Energy Sciences, under Contract No. DE-AC02-06CH11357. The beamline 34-ID-C was built with U.S. National Science Foundation grant DMR-9724294. Work performed at UCL was supported by EPSRC and ERC. Work at Brookhaven National Laboratory was supported by the U.S. Department of Energy, Office of Science, Office of Basic Energy Sciences, under Contract No. DE-SC0012704.
For the purpose of open access, the authors have applied a Creative Commons Attribution (CC BY) license to any Author Accepted Manuscript version arising.

\newpage

%% References
\bibliographystyle{sn-basic}
\bibliography{bibliography}

\end{document}